# Observation of the fractional quantum spin Hall effect in moiré MoTe$_2$

Kaifei Kang[1*], Bowen Shen[1], Yichen Qiu[2], Kenji Watanabe[3], Takashi Taniguchi[3], Jie Shan[1,2,4*] and Kin Fai Mak[1,2,4*]

[1] School of Applied and Engineering Physics, Cornell University, Ithaca, NY, USA
[2] Department of Physics, Cornell University, Ithaca, NY, USA
[3] National Institute for Materials Science, Tsukuba, Japan
[4] Kavli Institute at Cornell for Nanoscale Science, Ithaca, NY, USA

[*]Emails: kk726@cornell.edu; jie.shan@cornell.edu; kinfai.mak@cornell.edu

**Quantum spin Hall (QSH) insulators are two-dimensional electronic materials that have a bulk band gap like an ordinary insulator but have topologically protected pairs of edge modes of opposite chiralities [1-6]. To date, experimental studies have found only integer QSH insulators with counter-propagating up-spins and down-spins at each edge leading to a quantized conductance $G_0 = e^2/h$ (with $e$ and $h$ denoting the electron charge and Planck's constant, respectively) [7-14]. Here we report transport evidence of a fractional QSH insulator in 2.1-degree-twisted bilayer MoTe$_2$, which supports spin-$S_z$ conservation and flat spin-contrasting Chern bands [15,16]. At filling factor $\nu = 3$ of the moiré valence bands, each edge contributes a conductance $\frac{3}{2}G_0$ with zero anomalous Hall conductivity. The state is likely a time-reversal pair of the even-denominator 3/2-fractional Chern insulators. Further, at $\nu = 2, 4$ and $6$, we observe a single, double and triple QSH insulator with each edge contributing a conductance $G_0, 2G_0$ and $3G_0$, respectively. Our results open up the possibility of realizing time reversal symmetric non-abelian anyons and other unexpected topological phases in highly tunable moiré materials [17-19].**

**Main**

The emergence of two-dimensional moiré materials has opened a new platform to explore quantum phases of matter that arise from electronic topology and correlations [17-20]. In particular, twisted transition metal dichalcogenides (tTMDs) in the AA-stacking structure form a honeycomb moiré lattice with two sublattice sites residing in separate TMD layers [15,16,21]. The topmost moiré valence bands, originated from the K- and K'-valley states of the monolayer TMDs, are spin-split and spin-valley locked by a large Ising (out-of-plane) spin-orbit field, which defines the spin quantization axis [22]. The continuum model band structure calculations have shown that flat Chern bands with spin/valley-contrasting Chern numbers emerge in small-angle tTMDs as a result of skyrmion-like interlayer hopping [15]. Both integer Chern insulator at moiré band filling factor $\nu = 1$ and fractional Chern insulators at $\nu = 3/5$ and $2/3$ have been recently reported in tMoTe$_2$ under zero magnetic field [23-26].

In this work, we study electrical transport in 2.1-degree tMoTe$_2$ with moiré density $n_M \approx 1.26 \times 10^{12}$ cm$^{-2}$. The smaller twist angle compared to the reported studies [23-26] (twist angle around 3.5 degrees and moiré density around $3.5 \times 10^{12}$ cm$^{-2}$) further flattens and isolates the Chern bands [15,16,27-35] and allows us to observe interaction effects at high filling factors ($v \geq 2$). By combining local and nonlocal transport with different measurement configurations, we observe, in addition to the reported Chern insulator at $v = 1$, a QSH insulator with one, two and three pairs of helical edge modes and each pair contributing a conductance $G_0$ at $v = 2$, 4 and 6, respectively (more than one pair is allowed because of spin-$S_z$ conservation [36]). But most surprisingly, we observe a QSH insulator at $v = 3$ with each edge contributing a conductance $\frac{3}{2}G_0$, which can only be explained by charge fractionalization [4,37-42]. Charge fractionalization is of significant interest to both fundamental physics and fault-tolerant topological quantum computing [43]. It has been observed in the fractional quantum Hall states in high-mobility two-dimensional (2D) electrons subjected to a high magnetic field [44] and in fractionally filled Hofstadter and Chern bands in moiré materials [23-26,45-47]. But unlike all known cases, where time reversal symmetry breaking (either explicitly or spontaneously) is essential, the fractional QSH insulator observed here is time reversal invariant.

**Device design and characterization**
We employ the dual-gated device structure (Fig. 1a), in which the top and bottom gates independently control the moiré band filling factor, $v$, and the out-of-plane electric field, $E$. The latter tunes the interlayer or moiré sublattice potential difference [23-26]. To facilitate contact formation for hole transport in tMoTe$_2$, we employ platinum (Pt) electrodes, contact gates to heavily hole-dope the tMoTe$_2$ regions adjacent to the metal electrodes, and monolayer WSe$_2$ between Pt and tMoTe$_2$ as a tunnel barrier. In this design (Extended Data Fig. 1), a tunnel barrier occurs only at the metal-heavily doped tMoTe$_2$ junctions; the junctions between the tMoTe$_2$ contact and channel regions are nearly transparent (see below). The contact resistance arises nearly entirely from the metal junctions. We achieve contact resistances down to about 25 kΩ (per contact) at low temperatures and is nearly independent of the channel filling factor down to $v < 1$. On the other hand, it remains challenging to achieve uniform moiré devices at very small twist angles as in this study likely due to structural relaxation [19]. We choose to study a small region (about $2 \times 5$ μm$^2$) with relatively uniform moiré as shown in the optical image of Fig. 1b. The region of interest (shaded in red) with four low resistance contacts (1-4, shaded in grey) has a twist angle of 2.1 degrees. The other regions of the device do not show the moiré effects and are electrically disconnected from the region of interest. See Methods for details on device fabrication, twist angle calibration (Extended Data Fig. 2) and contact resistance characterization (Extended Data Fig. 1).

We perform electrical transport studies as described in Methods. Unless otherwise specified, all measurements are at lattice temperature $T = 20$ mK. Figure 1d,e show the device four-terminal longitudinal ($R_{xx}$) and transverse ($R_{xy}$) resistances as a function of $E$ and $v$. Here $R_{xx}$ and $R_{xy}$ are the field symmetric and anti-symmetric parts of the resistance measured at small out-of-plane magnetic fields $B_\perp = \pm 0.1$ T using the configuration illustrated in the inset of Fig. 1d. A line cut of Fig. 1d,e along $E \approx 0$ is shown in the upper panel of Fig. 1c. At $v = 1$, we observe a quantized $R_{xy} = \frac{h}{e^2}$ accompanied by a vanishing $R_{xx}$. This is an integer Chern insulator, consistent with earlier studies on larger twist angle tMoTe$_2$ (Ref. [23-26]). In addition to $v = 1$, we observe a series of states with nearly vanishing $R_{xx}$ at integer fillings up to about $\nu = 8$ but nearly no Hall response (see Fig. 2a and inset for more evidence of $R_{xy} \approx 0$). We focus on the most prominent ones at $\nu = 2, 3, 4$ and 6. The vanishing $R_{xy}$ and $R_{xx}$ suggests that these states are neither trivial band insulators nor Chern insulators.

The nature of these states changes abruptly above certain electric field, $E_c$, which is marked schematically by the black dashed lines in Fig. 1d,e. Specifically, the states at $\nu = 1$ and 2 turn into highly insulating states for $E > E_c$ while those at $\nu = 3, 4$ and 6 become metallic; the Hall response is nearly absent irrespective of fillings for $E > E_c$. An electric-field-induced topological phase transition for the $\nu = 1$ Chern insulator has been demonstrated in larger twist angle tMoTe$_2$ (Ref. [23-26]). The moiré bands in tMoTe$_2$ are topologically nontrivial only in the layer-hybridized regime for $E < E_c$ (Ref. [15,16,33]). The critical electric field $E_c$ generally increases with filling because higher electric fields are required by electrostatics to polarize more charges to one layer. Similar electrostatics phase diagram has been observed in other TMD moiré bilayers [19,48].

We further characterize the device nonlocal four-terminal resistance, $R_{NL}$, as a function of $E$ and $\nu$ in Fig. 1f using the measurement configuration illustrated in the inset. A line cut along $E \approx 0$ is shown in the lower panel of Fig. 1c. We note that the device channel is highly symmetric; $R_{NL}$ is nearly identical when the source-drain and voltage probe pairs are swapped (Extended Data Fig. 3). We observe a vanishing $R_{NL}$ at $\nu = 1$, which is consistent with chiral edge transport associated with the Chern insulator. But $R_{NL}$ exhibits prominent peaks at $\nu = 2, 3, 4$ and 6. Further, large $R_{NL}$ is present only in the layer-hybridized regime. It becomes negligible (~ 10 Ω or less) compared to $R_{xx}$ (~ 2 - 5 kΩ) for $E > E_c$. All these results indicate edge dominated transport at $\nu = 2, 3, 4$ and 6 in the layer-hybridized regime.

**Integer and fractional QSH insulators**
We perform comprehensive two-terminal resistance measurements (Figs. 2, 3) to verify edge channel transport at $\nu = 2, 3, 4$ and 6 and to determine contribution per edge to conductance [8,10]. As discussed above, two-terminal resistance is the sum of resistances in series including the channel resistance and contact resistance (Extended Data Fig. 1a). We calibrate the contact resistance for each measurement configuration as the difference between the two-terminal resistance and four-terminal resistance ($R_{xx}$) when the tMoTe$_2$ channel is heavily hole doped and has a negligible $R_{xx} < 200$ Ω (Methods and Extended Data Fig. 1c). Two-terminal channel resistance, $R_{2t}$, is obtained by subtracting a constant (i.e. doping-independent) contact resistance from the measured resistance. Figure 2a shows the filling dependence of $R_{2t}$ near $E = 0$ for the measurement configuration illustrated in the inset of Fig. 2b. Figure 2b shows the corresponding two-terminal conductance, $G_{2t} = 1/R_{2t}$. A small magnetic field $B_\perp = 0.3$ T is applied to fully polarize the $\nu = 1$ Chern insulator (inset of Fig. 2a), but has negligible effect on $R_{2t}$ or $G_{2t}$ (Extended Data Fig. 7). The quantized $R_{2t} = \frac{h}{e^2}$ or $G_{2t} = G_0$ plateau for the $\nu = 1$ Chern insulator state validates our procedure of contact resistance calibration.

Similar to the Chern insulator at $\nu = 1$, we observe nearly quantized plateaus of $R_{2t} = \frac{h}{\nu e^2}$ or $G_{2t} = \nu G_0$ around $\nu = 2$, 3, 4 and 6 (Fig. 2). The conductance plateaus are electric-field independent for small fields ($E < E_c$), whereas conductance in the compressible regions between two integer fillings is sensitive to $E$ (Fig. 2b). The electric-field dependence of $G_{2t}$ is illustrated in Fig. 3b-d for $\nu = 2$, 3, 4 and Extended Data Fig. 4 for $\nu = 1$, 6. We also include the result for three other measurement configurations. The color of the lines matches the color of the channel in the measurement schematics in Fig. 3a. An electric-field plateau is observed for all cases, and the electric-field span of the conductance plateaus agrees well with the layer-hybridized regime (between the arrows) identified in Fig. 1. Further, the value of the conductance plateaus is dependent on the measurement configuration for $\nu = 2$, 3, 4 and 6. It closely matches $\nu$, $\frac{3}{4}\nu$, $\frac{2}{3}\nu$ and $\frac{1}{2}\nu$ (in units of $G_0$), respectively, for the configuration from left to right in Fig. 3a. In contrast, the conductance plateau $G_{2t} \approx G_0$ at $\nu = 1$ is nearly independent of measurement configuration, which is consistent with chiral edge channel transport [10].

The transport behavior observed at $\nu = 2$, 3, 4 and 6 can be fully described as helical edge transport within the Landauer-Buttiker formalism (Methods). The helical edge states are equilibrated at each contact and are populated according to the chemical potential of the contact from which they emanate [8,10]. This leads to a quantized resistance per edge (that is, edge between two adjacent contacts). We can compute $R_{2t}$ or $G_{2t}$ for each measurement configuration using the equivalent circuit in Fig. 3a. The quantized resistance or conductance per edge is $\frac{2h}{\nu e^2}$ or $\frac{1}{2}\nu G_0$ to describe the results in Figs. 2, 3 (dashed lines). Helical edge transport also explains the observed $R_{xx}$, $R_{xy}$ and $R_{NL}$ at $\nu = 2$, 3, 4 and 6 in Fig. 1c: The measurement configuration for the four-terminal resistance in Fig. 1d is effectively a bridge circuit, which gives a vanishing $R_{xx}$ (whereas $R_{xx}$ for bulk transport does not vanish); The anomalous Hall response is absent because there is no net chirality; The four-terminal nonlocal configuration in Fig. 1f gives $R_{NL} = \frac{1}{2\nu}\frac{h}{e^2}$ (dashed lines in Fig. 1c) by counting the number of edges involved.

All the results support that the $\nu = 2$, 3, 4 and 6 states are QSH insulators in the layer-hybridized regime. Specifically, for even filling factor $\nu = 2$, 4 and 6, each edge contributes a conductance $G_0$, $2G_0$ and $3G_0$, respectively; these states are consistent with integer QSH insulators with one, two and three pairs of helical edge modes and each pair contributing a conductance $G_0$. For odd filling factor $\nu = 3$, each edge contributes a conductance $\frac{3}{2}G_0$, which can be explained only by charge fractionalization; the $\nu = 3$ state is consistent with a fractional QSH insulator [4,37-42].

**Temperature dependence**
We examine the energy scales of the QSH states by studying their temperature dependence. Figure 4a shows the filling factor dependence of $G_{2t}$ near $E = 0$ with the same measurement configuration as in Fig. 2. The curves correspond to different temperature ranging from 20 mK to 12 K, and are vertically displaced for clarity. The contact resistance is calibrated at each temperature (Extended Data Fig. 1d). As temperature increases, the conductance plateaus are gradually smeared out, and at high temperatures the conductance smoothly increases with filing factor. Specifically, the QSH plateaus at $\nu = 3$, 4 and 6 are smeared out above about 4, 10 and 6 K, respectively; the $\nu = 2$ QSH plateau remains visible above 12 K (as a reference, the temperature of the Chern insulator at $\nu = 1$ is about 6 K). We estimate the gap size for the fractional QSH insulator to be about 0.3 meV. Similar energy scales are observed from the temperature dependence of four-terminal $R_{NL}$ (Extended Data Fig. 5).

Figure 4b shows the electric-field dependence of $G_{2t}$ for the fractional QSH insulator at varying temperatures (additional data for the $\nu = 1$, 2, 4 states are included in Extended Data Fig. 5). The conductance is weakly temperature dependent in the layer-hybridized regime with $E < E_c$; in the layer-polarized regime with $E > E_c$, $G_{2t}$ decreases with increasing temperature, which is characteristic for a metal. The result suggests an electric-field-induced transition from a fractional QSH insulator to a Fermi liquid. Future studies based on Corbino geometry devices, which can access bulk transport in the fractional QSH insulator, are warranted to quantitatively study the quantum phase transition.

**Discussions and conclusions**
The observation of integer QSH insulators with a staircase-like dependence (Fig. 2 and 4) for the one, two and three pairs of helical edge states at $\nu = 2$, 4 and 6, respectively, reveals the presence of multiple Landau-level-like Chern bands in 2.1-degree tMoTe$_2$. Our result is compatible with first three moiré bands having Chern number +1 for the K valley and Chern number -1 for the K' valley (so that the number of helical edge state pairs adds up with increasing $\nu$). This sequence of Chern bands appears to be consistent with a recent band structure calculation [49]. An even number of helical edge state pairs is allowed here because they are protected by the spin-$S_z$ conservation symmetry [22,36]. The Ising-like spin alignment of the helical edge states is supported by the experimental observation of anisotropic magneto-transport in Extended Data Fig. 7. Whereas $G_{2t}$ is nearly independent of $B_\perp$ at $\nu = 2$, 3, 4 and 6, it decreases with in-plane magnetic field $B_\parallel$, which can induce spin mixing and backscattering [11-13].

The QSH insulator at $\nu = 3$ is of interaction origin; it occurs only in very small angle tMoTe$_2$ (Extended Data Fig. 6). The observed zero Hall conductivity excludes the possibility of the state being a $\nu = 2$ QSH insulator plus a $\nu = 1$ valley-polarized Chern insulator. The measured conductance, $\frac{3}{2}G_0$ per edge, can only be explained by charge fractionalization [4,37-42]. Further, the spin-$S_z$ conservation suggests that the state could be composed of two time-reversal-copies of the spin-polarized even-denominator-3/2 fractional Chern insulator. Such an assignment is compatible with the above proposed sequence of spin/valley-contrasting Chern bands with negligible spin/valley mixing. The $\nu = 3$ state is thus potentially a non-abelian fractional QSH insulator.

The observation of a fractional QSH insulator at $\nu = 3$ in tMoTe$_2$ leaves many open questions. One such question is the stability of the state. The fractional quantum Hall (FQH) state at 3/2 filling of the spin/valley-polarized Landau levels has been observed in in monolayer TMD under high magnetic fields [50]; and the 3/2-FQH state is more robust than the odd-denominator 4/3, 5/3-FQH states. This highlights the distinct electronic properties of the TMD-based and the GaAs-based 2D electron systems. In addition, insulating states at integer filling factors are generally more robust than those at fractional fillings in moiré materials [19]. These points may shed light on the stability of the $\nu = 3$ fractional QSH state. Future experimental and theoretical studies are required to fully understand the nature of this exotic state and competition with other symmetry breaking phases.

## Methods
### Device fabrication
Dual-gated devices of tMoTe$_2$ were fabricated using the tear-and-stack and layer-by-layer transfer method [51,52]. Figure 1a shows the device schematic. In short, flakes of monolayer MoTe$_2$, monolayer WSe$_2$, hexagonal boron nitride (hBN) and few-layer graphite were first exfoliated from bulk crystals onto Si/SiO$_2$ substrates and identified by their optical contrast. They were picked up using a PC (polycarbonate) thin film on PDMS (polydimethylsiloxane) in the following sequence: hBN, top gate (TG) graphite, TG hBN, part of the MoTe$_2$ monolayer, the rest of the MoTe$_2$ monolayer with a small twist angle, two WSe$_2$ monolayers as the tunnel barrier for the contacts, bottom gate (BG) hBN and BG graphite. The finished stack was released onto a Si/SiO$_2$ substrate with pre-patterned platinum (Pt) electrodes in the Hall bar geometry at a temperature of 180°C. The device channel is defined by the TG graphite electrode. The contact and split gates (CG and SG) were defined by standard electron beam lithography and evaporation of palladium. The split gate turns off the tMoTe$_2$ regions that are covered only by the bottom gate so that they do not contribute to transport. The tMoTe$_2$ channel is connected to the Pt electrodes through heavily hole-doped tMoTe$_2$ contact regions, which are controlled by the contact gates, to achieve low contact resistance (Extended Data Fig. 1a). The device was annealed at 200°C under high vacuum (< $10^{-5}$ mbar) for six hours to further improve the contacts. We have examined two devices of different twist angles in this study (Extended Data Fig. 6).

### Electrical measurements
Electrical transport measurements were performed in a Bluefors LD250 dilution refrigerator equipped with a 12 T superconducting magnet. Low-frequency (11.77 Hz) lock-in techniques were used to measure the sample resistance under a constant bias voltage (0.3 mV), which excites a bias current of less than 10 nA to avoid sample heating and/or high bias effects. Voltage pre-amplifiers with large input impedance (100 MΩ) were used in the four-terminal resistance measurements.

### Twist angle calibration
We calibrated the twist angle of MoTe$_2$ moiré based the Landau levels originated from $\nu = 2$ under high magnetic fields (Extended Data Fig. 2). Landau levels with index $\nu_{LL} = 1$, 2 and 3 are identified from nearly quantized $R_{xy}$ and $R_{xx}$ minimum. They are consistent with Landau levels being fully spin- and valley-polarized [50], for which the density difference between two successive Landau levels is $B_\perp/\Phi_0$, where $\Phi_0 = h/e$ denotes the magnetic flux quantum. We determine the moiré density $n_M$ based on the known moiré filling factor $\nu$. Using the moiré density we determine the twist angle $\theta = a\sqrt{\frac{\sqrt{3}}{2}n_M}$, where $a \approx 3.5$ Å is the MoTe$_2$ lattice constant. Extended Data Fig. 2d shows the twist angle calibrated at different magnetic fields between 9 T and 11 T. It is independent of field. The mean is 2.10 degrees with an uncertainty of about $\pm 0.05$ degrees. The corresponding moiré density is $n_M \approx 1.26 \times 10^{12}$ cm$^{-2}$.

**Contact resistance characterization**

Extended Data Fig. 1a shows a schematic of the electrical contacts to the device channel. The metal (Pt) electrodes are connected to the tMoTe$_2$ channel through a region of heavily hole-doped tMoTe$_2$. A tunnel barrier occurs at the metal-to-heavily doped tMoTe$_2$ junctions, but the junctions between the heavily-doped tMoTe$_2$ contact and the lightly-doped tMoTe$_2$ channel regions are nearly transparent (as evidenced by the quantized $R_{2t}$ and $R_{xy}$ in Fig. 2a) [53]. The contact resistance arises nearly entirely from the metal junctions. The two-terminal resistance is the sum of resistances in series including the channel resistance and the two contact resistances [54].

Extended Data Fig. 1b illustrates the contact characterization for the two-terminal configuration shown in the inset. The results for other two-terminal configurations are similar. The raw two-terminal resistance $R_{2t}$ (measured near $E \approx 0$ at 20 mK) decreases with increasing filling factor. The different curves are for different contact gate voltages. They are nearly identical for filling factor down to $\nu < 1$ except a constant resistance offset; a more negative contact gate voltage corresponds to a smaller $R_{2t}$. This is consistent with filling independent contact resistances, the value of which is only controlled by the doping density in the contact region. A more negative contact gate voltage increases the hole doping density in the contact region and decreases the contact resistance. The result is fully consistent with the contact design in Extended Data Fig. 1a.

To calibrate the value of the contact resistance, we compare the raw two-terminal resistance with the four-terminal resistance $R_{xx}$ under the same conditions when the channel is heavily doped (Extended Data Fig. 1c). The four-terminal resistance is below 200 Ω for $\nu > 10$; and the value is nearly temperature independent (inset). We subtract a constant resistance (the contact resistance) from the raw two-terminal resistance to match $R_{xx}$ at large $\nu$. Figure 2a shows $R_{2t}$ for the measurement configuration given in the inset of Fig. 2b after subtraction of 26.8 kΩ. It shows $R_{2t} \approx \frac{h}{e^2}$ at $\nu = 1$, which agrees well with the expected quantized value for a Chern insulator (supported by the quantized $R_{xy}$ and vanishing $R_{xx}$). The result supports the contact model in Extended Data Fig. 1a and the validity of the calibration procedure of the contact resistance. The quantized $R_{xy}$ and $R_{2t}$ at $\nu = 1$, in accord with the Landauer-Buttiker formalism with transparent contacts, also support that the junctions between heavily-doped and lightly-doped tMoTe$_2$ are nearly transparent [8,10,53]. Extended Data Fig. 1d illustrates the temperature dependence of the calibrated contact resistance (for one contact). The weak temperature dependence indicates that the contacts are in the tunnel contact limit. The contact design here involving a tunnel barrier yields a contact resistance that is smaller than the reported values in earlier studies of tMoTe$_2$ (about 100 kΩ) [24,25].

**Landauer-Buttiker analysis of ballistic edge transport**

The low-temperature ballistic edge transport can be modeled using the Landauer-Buttiker analysis. Unlike the chiral edge states, the helical edge states can be equilibrated at the electrical contact and are therefore populated according to the chemical potential of the contact from which they emanate [8,10]. This leads to the equivalent circuits in Fig. 3a with a quantized resistance $\frac{2h}{\nu e^2}$ per edge (i.e. between two adjacent contacts) for the QSH insulators at $\nu = 2, 3, 4$ and 6. Note that the edge between two shorted electrodes, which share the same chemical potential, does not contribute any resistance. We can compute the two-terminal channel resistance, $R_{2t} = \frac{1}{\nu}, \frac{4}{3\nu}, \frac{3}{2\nu}, \frac{2}{\nu}$ (in units of $\frac{h}{e^2}$), for the measurement configurations from left to right in Fig. 3a. In contrast, $R_{2t}$ for the $\nu = 1$ Chern insulator with a chiral edge state is expected to be independent of the measurement configuration if no equilibration with the contacts occurs. Figure 3b-d and Extended Data Fig. 4 show good agreement between our experiment and the above analysis.

Similarly, we can compute the four-terminal nonlocal resistance $R_{NL}$ with the measurement configuration shown in the inset of Fig. 1f for the QSH insulators. The nonlocal resistance is determined by $R_{NL} = \frac{V_{NL}}{I}$, where $I$ is the total two-terminal bias current and $V_{NL}$ is the voltage drop at one edge. We obtain $R_{NL} = \frac{1}{2\nu}\frac{h}{e^2}$ for $\nu = 2, 3, 4$ and 6 (dashed lines in Fig. 1c). The experimental result is in good agreement with the quantized value at $\nu = 2$, 10-20 % smaller at $\nu = 3$ and 4, and substantially smaller at $\nu = 6$. The discrepancies are likely caused by bulk conduction, particularly for the smaller gap states at higher filling factors. (Note that $R_{NL}$ is exponentially suppressed [55] for the bulk-dominate transport for $E > E_c$.) Using the same circuit model, we can obtain $R_{xx} = 0$ for the QSH insulators in the measurement configuration of Fig. 1d, which is effectively a bridge circuit for the helical edge resistors. Again, the results (Fig. 1c) are in good agreement with the Landauer-Buttiker analysis.

**Acknowledgements**
We thank Chaoming Jian, Allan MacDonald, Charles Kane, Liang Fu, Andrei Bernevig, Nicolas Regnault, Jiabin Yu, Trithep Devakul, Aiden Reddy and Eun-Ah Kim for many helpful discussions.

Figures

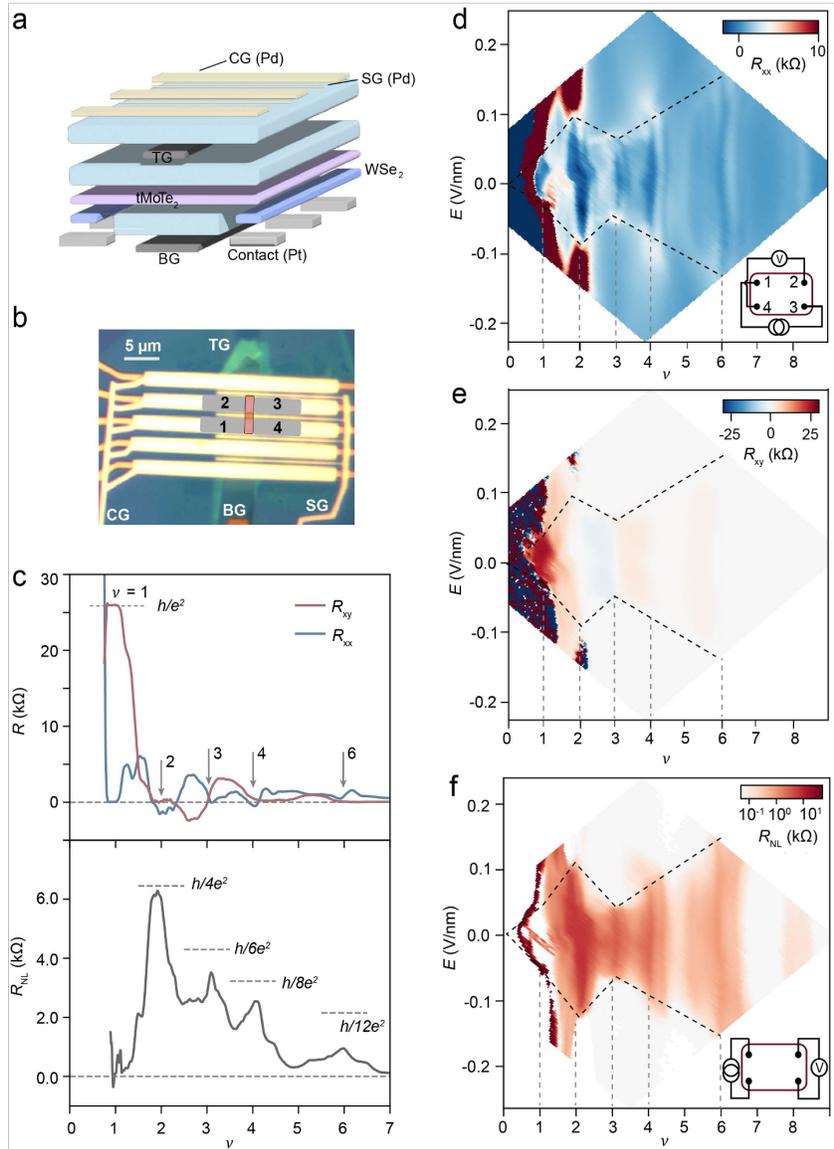

**Figure 1 | Moiré MoTe$_2$ device and characterization. a,** Schematic of dual-gated tMoTe$_2$ devices. Both top gate (TG) and bottom gate (BG) are made of few-layer graphite contacts (dark grey) and hBN dielectrics (light blue). Monolayer WSe$_2$ (blue) is a tunnel barrier between the Pt contact electrodes (grey) and tMoTe$_2$ (purple). Palladium (yellow) is used as contact gates (CG) and split gates (SG). **b,** Optical micrograph of the 2.1-degree tMoTe$_2$ device with region of interest shaded in red and four contact electrodes (1-4) shaded in grey. **c,** Line cut of **d-f** at $E \approx 0$. The dashed lines denote the expected resistance around $\nu = 1, 2, 3, 4$ and 6. **d-f,** Four-terminal resistance $R_{xx}$ (**d**), Hall resistance $R_{xy}$ (**e**) and nonlocal resistance $R_{NL}$ (**f**) versus vertical electric field $E$ and filling factor $\nu$ at 20 mK. The measurement configurations for **d**, **e** and for **f** are shown in the inset of **d** and **f**, respectively. $R_{xx}$ and $R_{xy}$ are the field symmetric and anti-symmetric parts of the measured resistance at $B_\perp = \pm 0.1$ T; $R_{NL}$ is measured under 0.1 T. The black dashed lines are guides to the eye of the layer-hybridized regime; the grey dashed lines denote filling factor $\nu = 1, 2, 3, 4$ and 6.

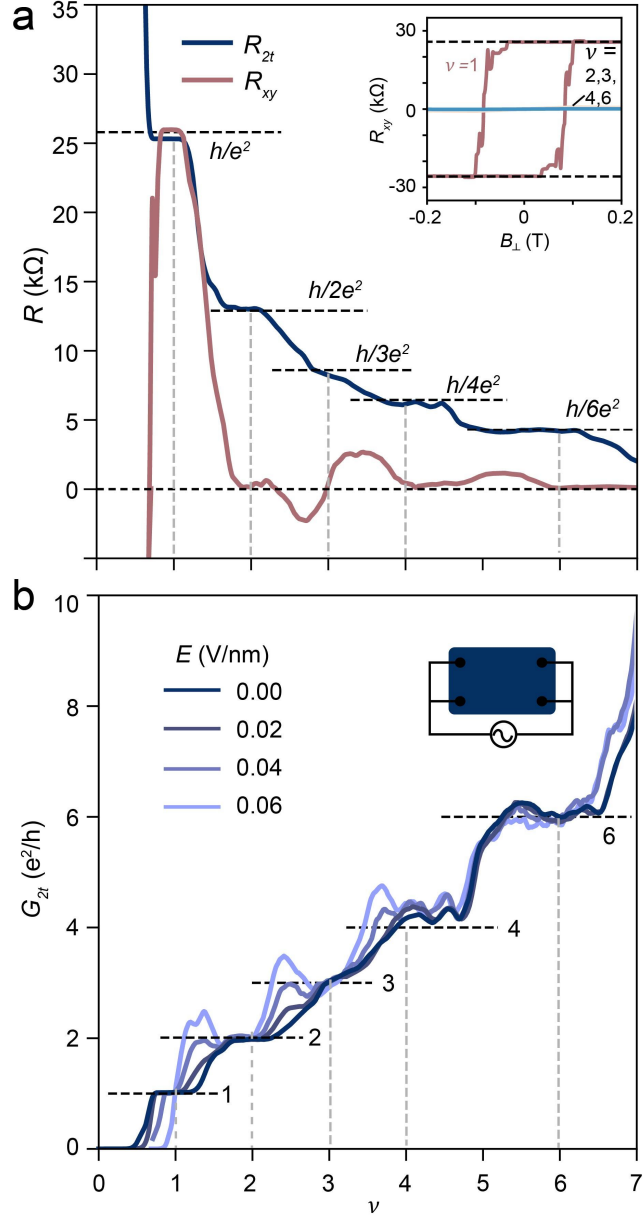

**Figure 2 | Quantized two-terminal transport. a,** Filling factor dependence of two-terminal resistance ($R_{2t}$) and Hall resistance ($R_{xy}$) at $E \approx 0$ and 20 mK. Inset: $R_{xy}$ shows a magnetic hysteresis and a quantized Hall resistance $\frac{h}{e^2}$ at zero magnetic field for $\nu = 1$; the Hall resistance is nearly zero for $\nu = 2, 3, 4$ and $6$. **b,** Filling factor dependence of two-terminal conductance, $G_{2t} = 1/R_{2t}$ (in units of $\frac{e^2}{h}$), at representative electric fields in the layer-hybridized regime. The two-terminal measurement configuration is illustrated in the inset of **b**. A constant contact resistance 26.8 kΩ is subtracted from the measured two-terminal resistance to obtain $R_{2t}$ (see Methods for details). Magnetic field $B_\perp = 0.3$ T is applied to fully polarize the $\nu = 1$ Chern insulator state. The horizontal dashed lines in **a,b** are the expected resistance/conductance around $\nu = 1, 2, 3, 4$ and $6$ (vertical dashed lines).

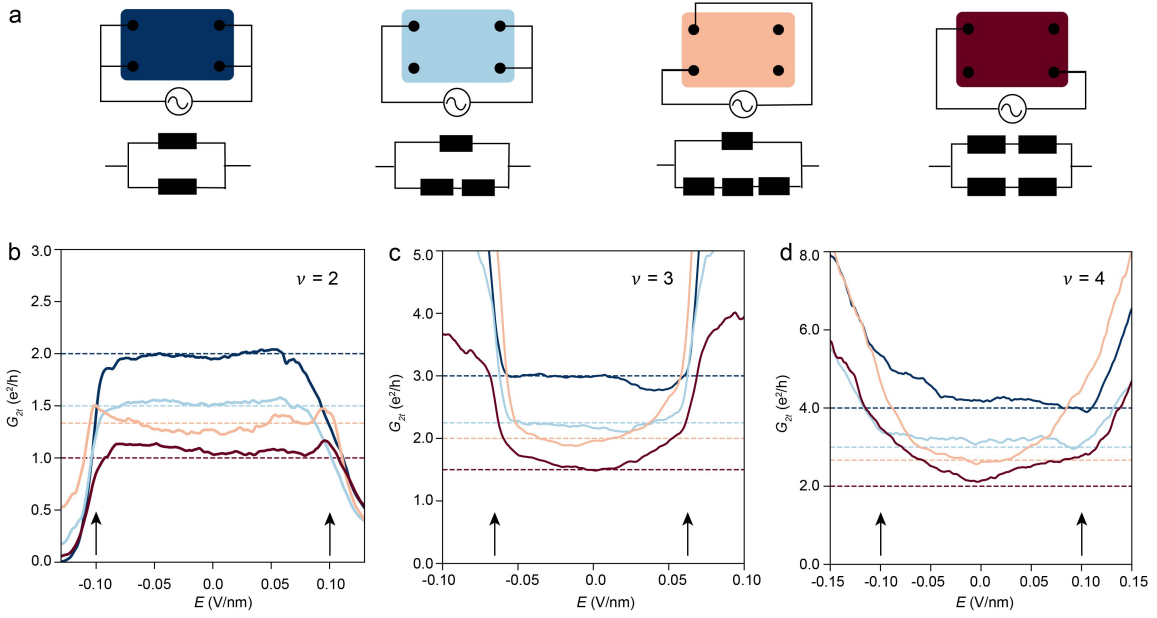

**Figure 3 | Two-terminal nonlocal transport. a,** Four two-terminal nonlocal measurement configurations (upper) and equivalent circuits for helical edge mode transport (lower). The shaded rectangles represent tMoTe$_2$ with four contacts (black dots). Unconnected dots are floated contacts. The equivalent resistor associated with helical edge transport is $\frac{2h}{\nu e^2}$ per edge for $\nu = 2$, 3 and 4. **b-d,** Electric-field dependence of $G_{2t}$ (in units of $\frac{e^2}{h}$) for different measurement configurations around $\nu = 2$ (**b**), $\nu = 3$ (**c**) and $\nu = 4$ (**d**) at 20 mK. The horizontal dashed lines denote the expected conductance $\nu$, $\frac{3}{4}\nu$, $\frac{2}{3}\nu$ and $\frac{1}{2}\nu$ in descending order from the Landauer-Büttiker analysis. The color of the lines matches the color of the channels in **a**. The layer-polarized regime is between the two arrows for each filling factor.

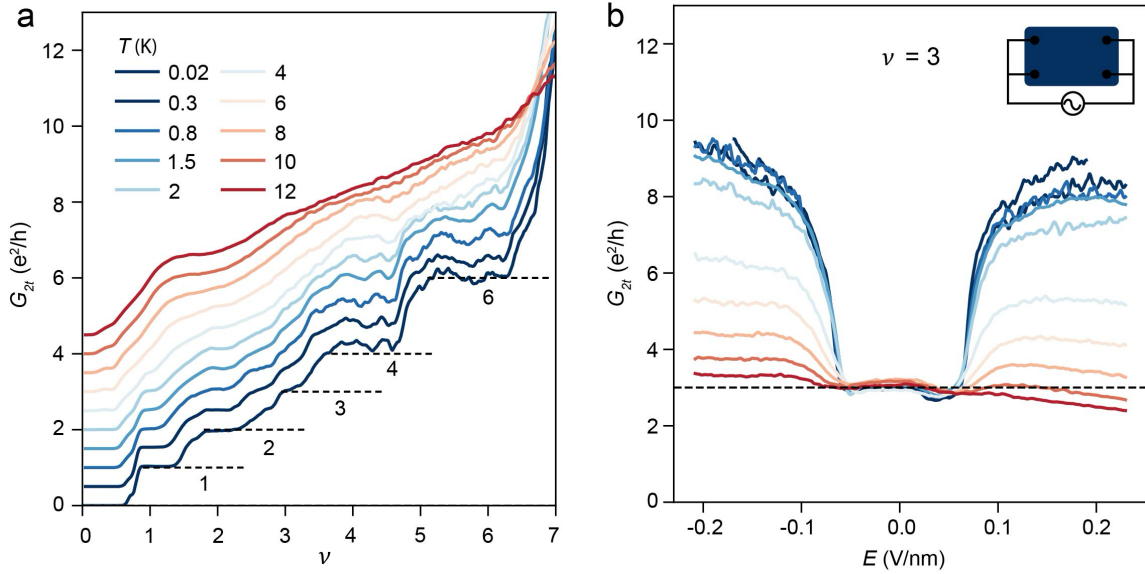

**Figure 4 | Topological phase transition and thermal excitations. a,** Filling factor dependence of $G_{2t}$ (in units of $\frac{e^2}{h}$, measured with the configuration in the inset of **b**) at $E \approx 0$ and temperature ranging from 0.02 K to 12 K. The curves at different temperatures are vertically displaced by 0.5 unit for clarity. **b,** Electric-field dependence of $G_{2t}$ at $\nu = 3$ and varying temperatures. A QSH insulator-to-metal transition is observed with increasing electric field at critical field $E_c \approx \pm 70$ mV/nm. The horizontal dashed lines in **a,b** denote the expected conductance $\nu \frac{e^2}{h}$ around $\nu = 1, 2, 3, 4$ and 6. Magnetic field $B_\perp = 0.3$ T is applied to fully polarize the $\nu = 1$ Chern insulator state.

**Extended Data Figures**

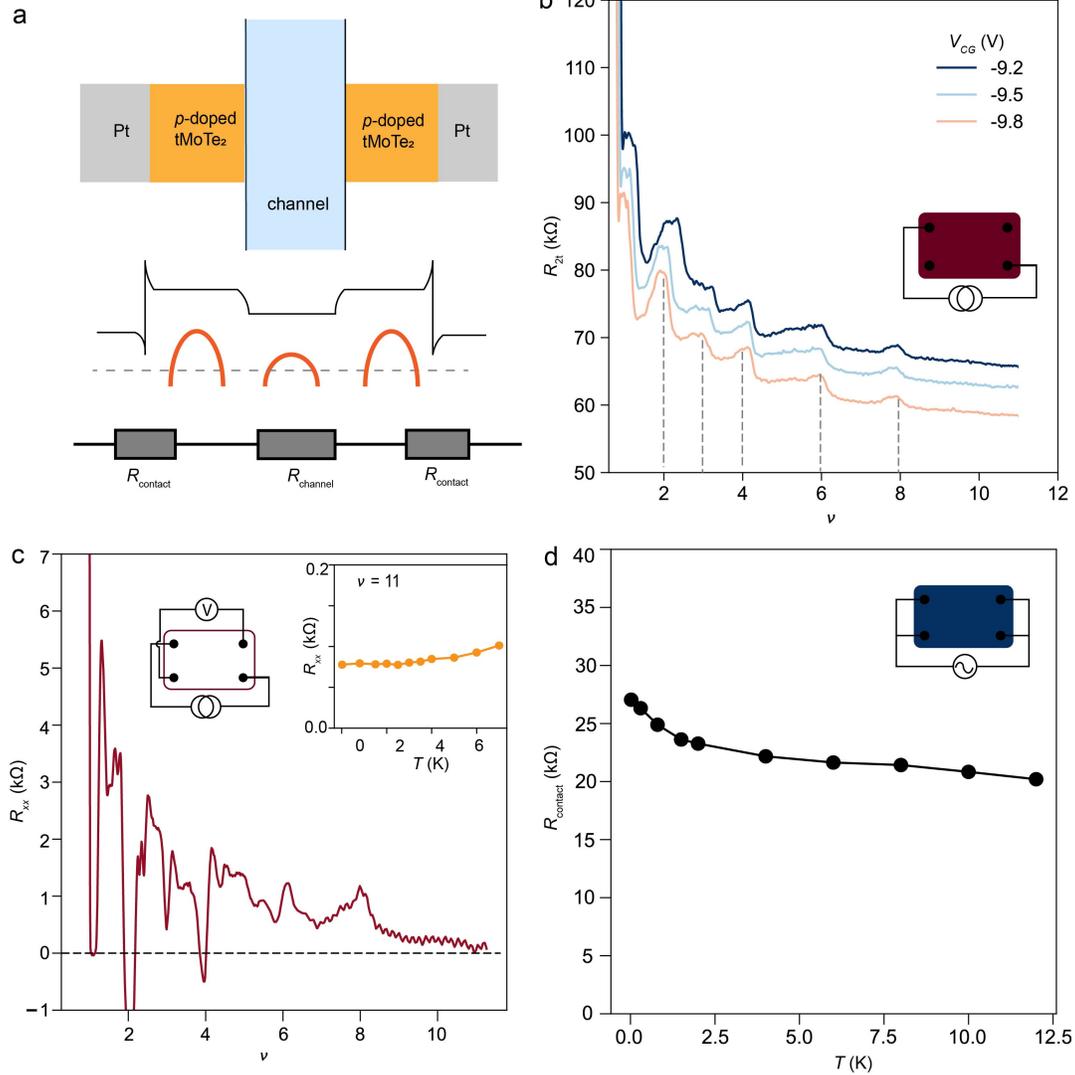

**Extended Data Figure 1 | Contact resistance characterization. a,** Top: schematic of the device contact and channel. The Pt electrodes (grey) are contacted to the tMoTe$_2$ channel (blue) through heavily hole-doped tMoTe$_2$ contact regions (orange). Middle: schematic valence band alignment of the tMoTe$_2$ contact and channel regions (solid red lines). Dashed line: Fermi level; solid black line: spatial variation of the work function. Tunnel barriers are at the Pt junctions, but the heavily-to-lightly doped tMoTe$_2$ junctions are transparent. Bottom: equivalent circuit model for two-terminal measurements with $R_{\text{channel}}$ and $R_{\text{contact}}$ denoting the channel resistance and the contact resistance at the Pt junction, respectively. **b,** Filling factor dependence of two-terminal resistance $R_{2t}$ at different contact gate voltages ($E \approx 0$, $T = 20$ mK and $B_\perp = 0.1$ T). Inset: measurement configuration. $R_{2t}$ at the highest fillings is approximately the sum of resistances in series for two contacts. **c,** Filling factor dependence of four-terminal resistance $R_{xx}$ measured with the configuration in the left inset. The resistance drops to about 100 Ω at $\nu = 11$, which is nearly temperature independent (right inset). **d,** Temperature dependence of contact resistance (for one contact) calibrated for the two-terminal configuration in the inset.

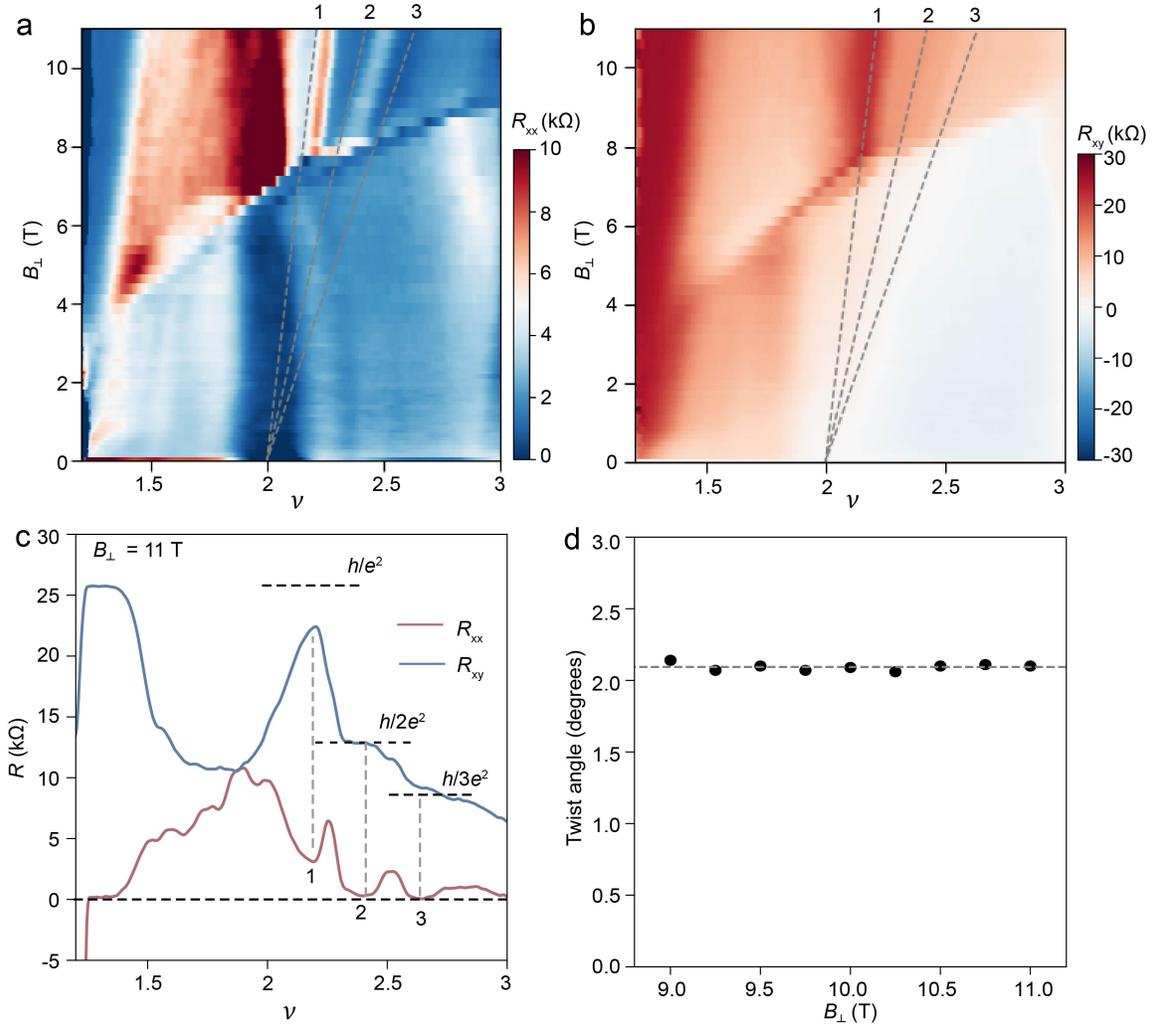

**Extended Data Figure 2 | Twist angle calibration. a,b,** Four-terminal resistance $R_{xx}$ (**a**) and Hall resistance $R_{xy}$ (**b**) versus out-of-plane magnetic field $B_\perp$ and filling factor $\nu$ at $E = 0.06$ V/nm and 20 mK. The measurement configuration is shown in the inset of Fig. 1d. Dashed lines denote Landau level $\nu_{LL} = 1$, 2 and 3 emerging from $\nu = 2$ above $B_\perp \approx 7 - 8$ T. The dispersing state at low filling factors is the $\nu = 1$ Chern insulator. **c,** Line cut of **a,b** at $B_\perp = 11$ T. The vertical dashed lines mark Landau level $\nu_{LL} = 1$, 2 and 3 with nearly quantized $R_{xy}$ and $R_{xx}$ minimum; the horizontal dashed lines denote the expected quantized value of $R_{xy} = \frac{h}{\nu_{LL} e^2}$. **d,** Twist angle calibrated from Landau level spacing at $B_\perp$ ranging from 9 T to 11 T. No field dependence is observed. The mean of the twist angle is 2.10 degrees (dashed line) and the uncertainty is about $\pm 0.05$ degrees.

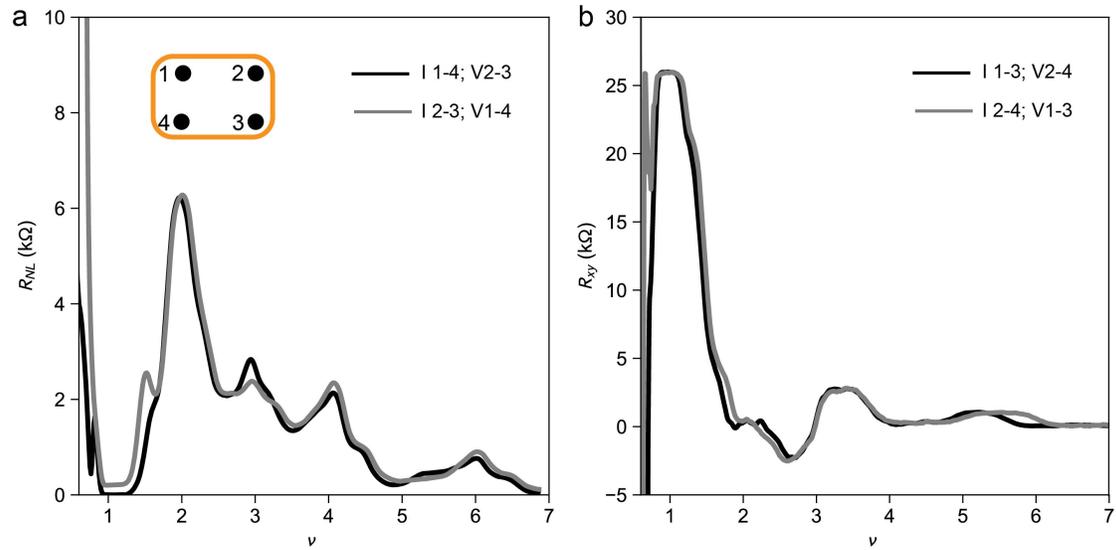

**Extended Data Figure 3 | Symmetric device channel. a,b,** Filling factor dependence of four-terminal nonlocal resistance $R_{NL}$ (**a**, $B_\perp = 0.1$ T) and Hall resistance $R_{xy}$ (**b**, $B_\perp = 0.2$ T) at $E \approx 0$ and $T = 20$ mK. The inset shows the device channel with contact 1-4. I: current bias; V: voltage probe. The two curves in each panel correspond to swapped source-drain and voltage probe pairs. The nearly identical results demonstrate a highly symmetric two-dimensional channel.

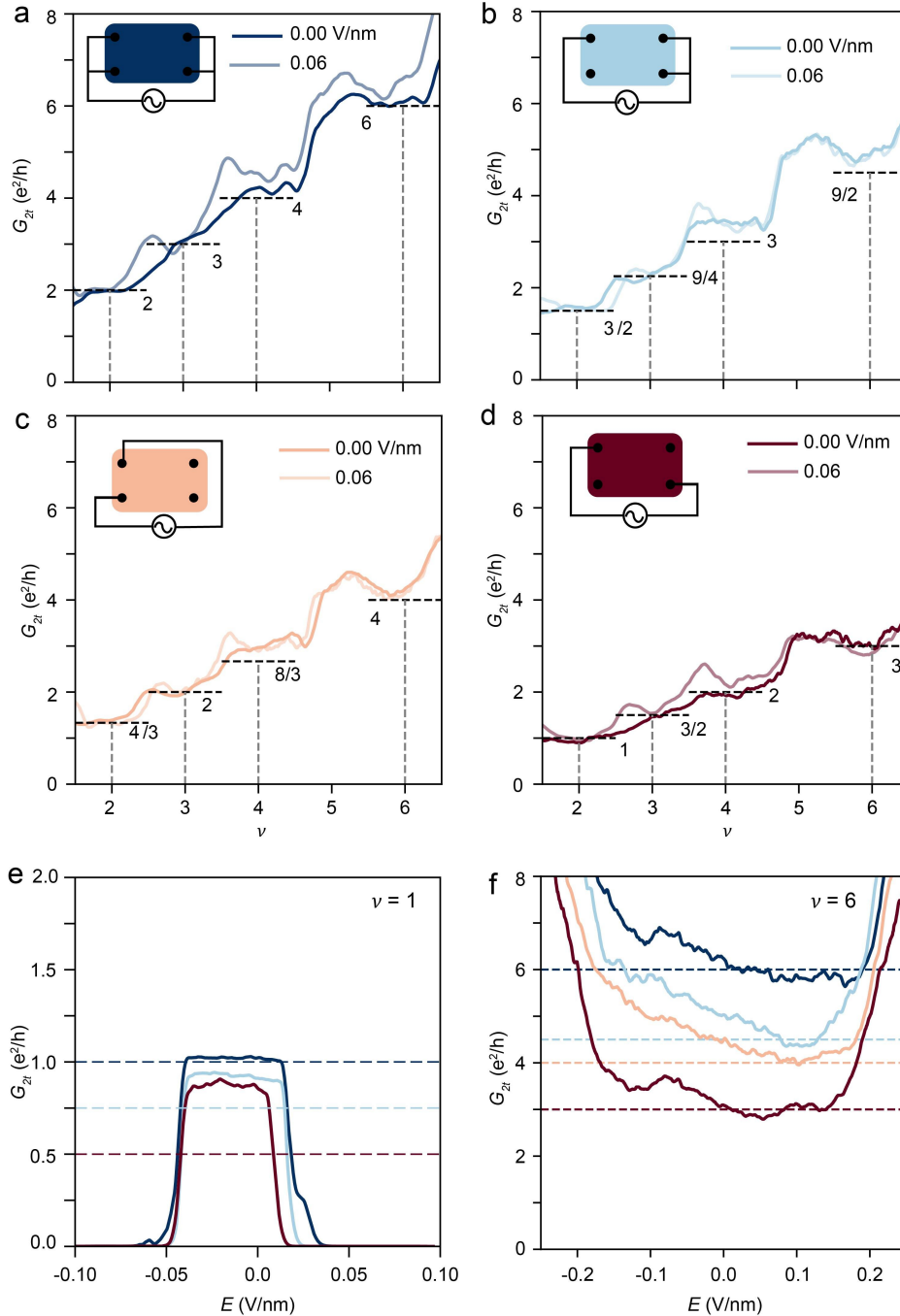

**Extended Data Figure 4 | Additional nonlocal two-terminal transport data. a-d,** Filling factor dependence of two-terminal conductance $G_{2t}$ (in units of $\frac{e^2}{h}$) at two electric fields inside the layer-hybridized regime. Inset: measurement configuration. The dashed lines denote the expected quantized values of $G_{2t}$ at $\nu =$ 2, 3, 4 and 6. **e,f,** Electric-field dependence of $G_{2t}$ (in units of $\frac{e^2}{h}$) at $\nu = 1$ (**e**) and $\nu = 6$ (**f**) at 20 mK. The horizontal dashed lines denote the expected conductance $\nu, \frac{3}{4}\nu, \frac{2}{3}\nu$ and $\frac{1}{2}\nu$ in descending order. The color of the lines matches the color of the channels in **a-d**. Magnetic field 0.3 T is applied to fully polarize the $\nu = 1$ Chern insulator.

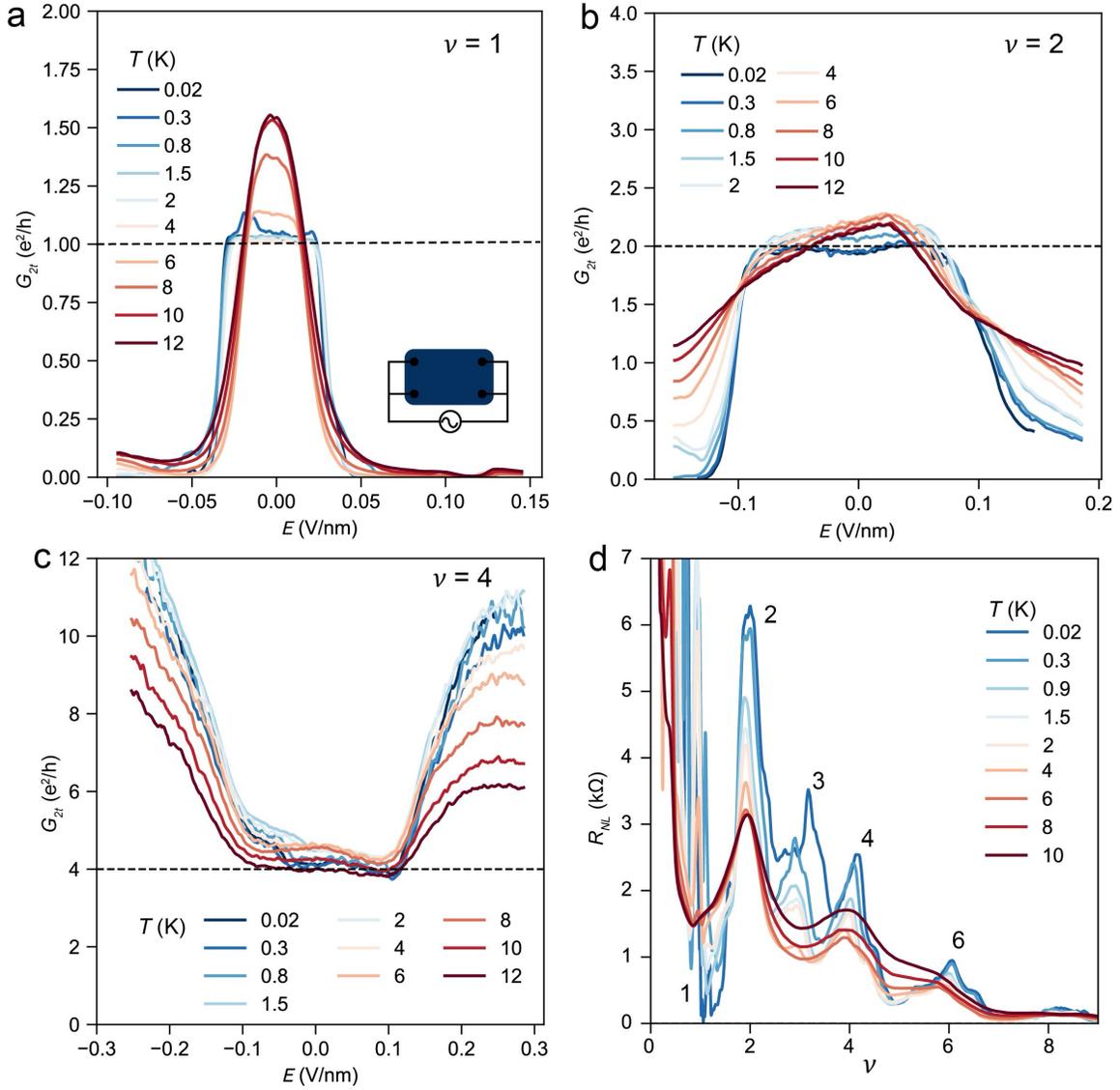

**Extended Data Figure 5 | Additional temperature dependence data. a-c,** Electric-field dependence of two-terminal conductance, $G_{2t}$ (in units of $\frac{e^2}{h}$), at varying temperatures for $\nu = 1$ (**a**), $\nu = 2$ (**b**) and $\nu = 4$ (**c**). The horizontal dashed lines denote the quantized value $G_{2t} = \frac{\nu e^2}{h}$. The measurement configuration is shown in the inset of **a**. **d,** Filling factor dependence of four-terminal nonlocal resistance $R_{NL}$ at $E \approx 0$ and varying temperatures. $R_{NL}$ at $\nu = 2$, 3, 4 and 6 decreases with increasing temperature as bulk conduction becomes more important.

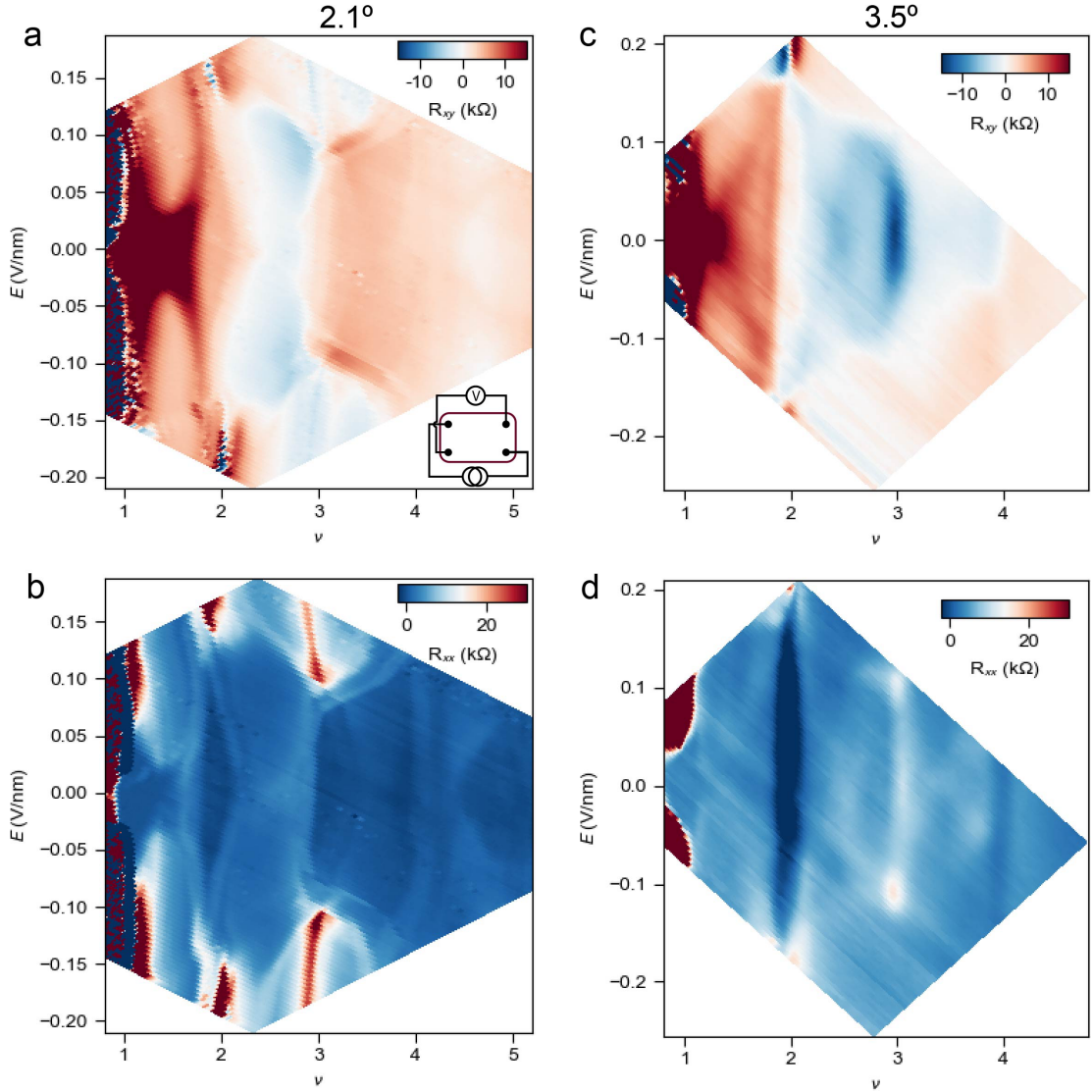

**Extended Data Figure 6 | Twist angle effects. a,b,** Four-terminal Hall resistance $R_{xy}$ (**a**) and longitudinal resistance $R_{xx}$ (**b**) versus out-of-plane electric field and filling factor for 2.1-degree tMoTe$_2$ device at $B_\perp = 5$ T and 20 mK. The measurement configuration is shown in the inset of **a**. Both $R_{xx}$ and $R_{xy}$ are small near $v = 3$ in the layer-hybridized regime even under high magnetic fields. **c,d,** Same as **a,b** for 3.5-degree tMoTe$_2$ device at $B_\perp = 4$ T and 1.6 K. A large negative $R_{xy}$ is accompanied by a weak $R_{xx}$ dip near $v = 3$ in the layer-hybridized regime (even though at higher temperature and lower magnetic field). The large Hall response suggests a valley-polarized state (rather than a fractional QSH insulator) at $v = 3$.

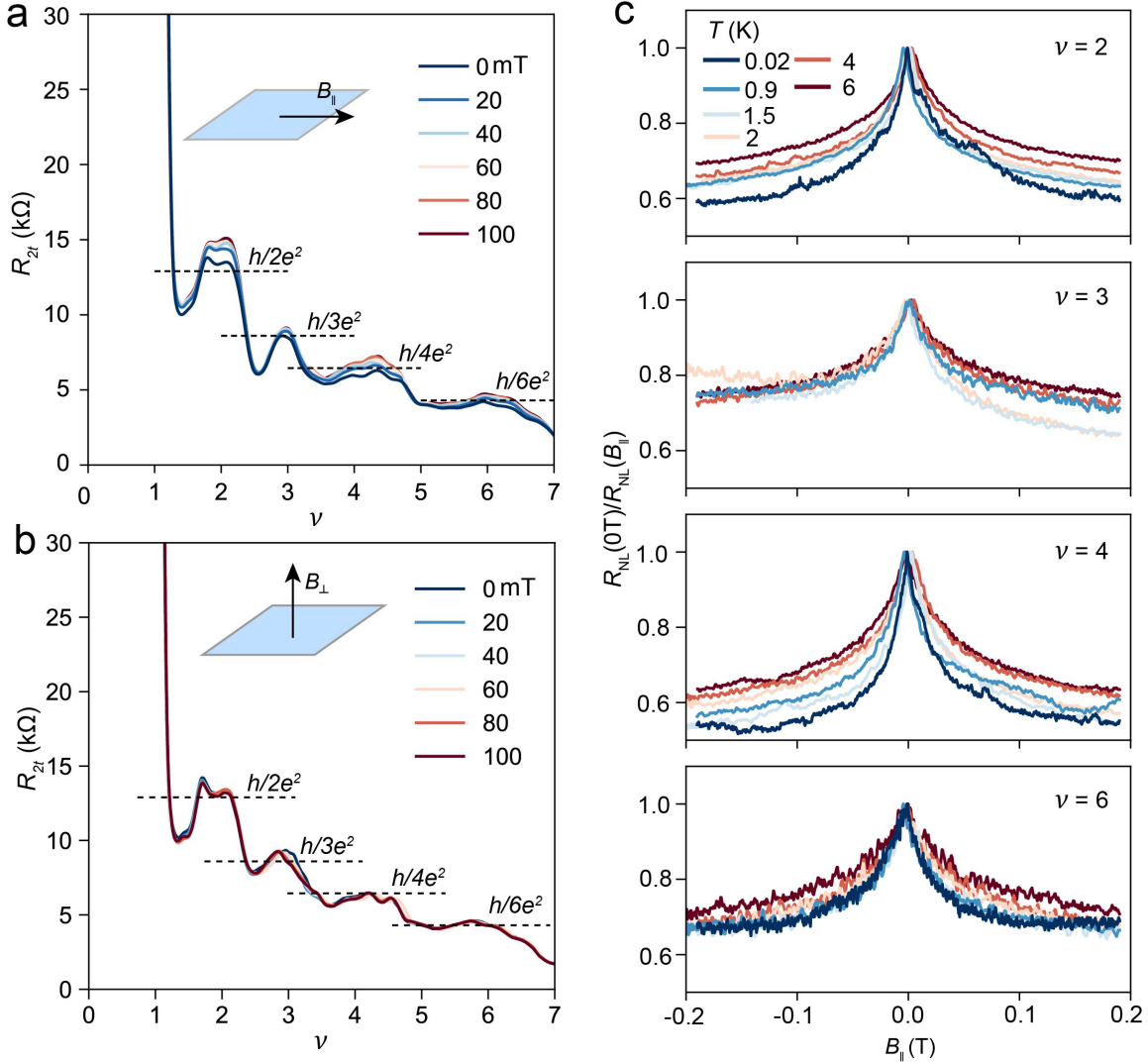

**Extended Data Figure 7 | Anisotropic magneto-response. a,b,** Filling factor dependence of two-terminal resistance $R_{2t}$ at varying in-plane (**a**) and out-of-plane (**b**) magnetic fields ($E = 0.06$ V/nm and $T = 20$ mK). Whereas $R_{2t}$ is nearly independent of $B_\perp$, it increases with $B_\parallel$. The horizontal dashed lines denote the quantized value $R_{2t} = \frac{h}{\nu e^2}$ at $\nu = 2, 3, 4$ and $6$. **c,** In-plane magnetic field dependence of $\frac{R_{NL}(B_\parallel = 0T)}{R_{NL}(B_\parallel)}$ at varying temperatures for $\nu = 2, 3, 4$ and $6$, where $R_{NL}$ is the nonlocal four-terminal resistance. At all filling factors the cusp at zero magnetic field is broadened as temperature increases. The anisotropic magneto-response supports that the helical edge states carry Ising spins. Because of spin-$S_z$ conservation, the edge states are immune to $B_\perp$, but are susceptible to gap opening and spin mixing under $B_\parallel$, thus increasing $R_{2t}$ and $R_{NL}$. The anisotropic magneto-response also contrasts with the expected bulk-dominant transport, which would show exactly the opposite magnetic field dependence, namely, strong out-of-plane but negligible in-plane magnetic field dependence.